\documentclass[aps,prb,groupedaddress,showpacs]{revtex4}
\usepackage{graphicx}

\begin{document}

\title{
Terahertz Response of Optically Excited Semiconductors}

\author{M. Kira}
\author{W. Hoyer}
\author{S. W. Koch}
\affiliation{Fachbereich Physik, Philipps Universit\"at, 
Renthof 5, 35032 Marburg/Germany}

\date{\today}

\begin{abstract}
The induced terahertz response of semiconductor systems is investigated
with a microscopic theory. In agreement with recent terahertz experiments, 
the developed theory fully explains the ultrafast build up of the plasmon
resonance and the slow formation of incoherent excitonic populations.
For incoherent conditions, it is shown that a terahertz field exclusively
probes the correlated electron-hole pairs via a symmetry breaking between
many-body correlations with even and odd functional form. 
\end{abstract}


\maketitle

\noindent

The light-matter coupling of semiconductor systems is usually 
mediated by the optical polarization generated by band-to-band 
excitations. This polarization --- i.e.~the induced coherence between the 
optically coupled valence and conduction bands --- is influenced and 
modified by already existing or excitation generated populations
of quasiparticles. However, since optical methods probing band-to-band
transitions cannot directly measure populations, such approaches provide
only indirect information about the characteristics of the semiconductor
excitations. 
Usually it is therefore very difficult, if not impossible,
to unambiguously attribute the observed changes in the interband
optical response to their genuine origin.
Hence, to learn about the true nature of optically
generated excitations and quasi-particles, one needs 
supplementary information from other methods. With this
respect, recent experimental 
efforts\cite{Groeneveld:94,Cerne:96,Beard:00,Huber:01,Kaindl:03} 
have extended semiconductor optics toward the regime where
transitions between quasi-particle states can be probed directly with far
infrared fields at terahertz (THz) frequencies which are orders of 
magnitude lower than the usual band-to-band transitions. 
Especially, the combination of interband optical excitation
and intraband measurement of the induced THz absorption 
allowed the experimentalists to discuss 
seemingly different questions such as
the ultrafast build-up of plasma screening\cite{Huber:01} and
the formation of excitonic correlations\cite{Kaindl:03}.

To provide a fundamental basis for the analysis of the semiconductor
THz response, we develop in this Letter a 
comprehensive microscopic theory that
consistently includes Coulombic many-body interactions,
coupling to optical and THz fields as well as lattice vibrations.
The developed theory is applied to 
compute the time-resolved THz response after 
optical interband excitation and as special cases, we investigate the 
build up of the plasmon resonance and 
the gradual formation of exciton population transitions.
Our results demonstrate that the incoherent
THz response can be attributed to transitions between
specific many-body states of the correlated electron-hole pairs,
establishing this technique as uniquely qualified to
identify the genuine nature and dynamics
of these quasi-particle excitations.

In general, the optical response of semiconductors to a classical transversal
electric field $E(z,t)$ can be solved from Maxwell's wave equation which 
couples the light-field to the (optical) interband polarization $P$
and to the (THz) intraband current $J$. We consider planar structures 
which can be either a quantum well (QW) or a set of identical quantum 
wires aligned periodically into a plane with a wire distance much
less than the wavelength. In order to compute the dynamic semiconductor 
response microscopically, we set up the equations of motion for 
the elementary semiconductor
quasi-particle excitations consisting of carriers, 
phonons, and photons in the formalism of second quantization.
The interactions are modeled by the standard Coulomb coupling
and the minimal-substitution Hamiltonian.\cite{Haug:94,Cohen:89} 
The general properties of a two-band 
electron-hole system are described by the Fermionic operators 
$a_{{\bf k},c(v)}$ and $a^\dagger_{{\bf k},c(v)}$
where, e.g., $a^\dagger_{{\bf k},c(v)}$ creates an electron  
with carrier momentum ${\bf k}$ in the conduction (valence) band.
Such a system has 
$P = \frac{1}{\cal S} \sum_{{\bf k}} d^{\star}_{\rm{cv}} P_{\bf k} + {\rm c.c.}$
and 
$J = \frac{1}{\cal S} \sum_{{\bf k},\lambda} 
\left[j_\lambda({\bf k}) - e^2 A/m_0\right] 
f^{\lambda}_{\bf k}$
with the normalization area ${\cal S}$,
microscopic polarization
$P_{\bf k} \equiv \langle a^\dagger_{{\bf k},v} a_{{\bf k},c} \rangle$,
electron 
$f^{e}_{\bf k} \equiv \langle a^\dagger_{{\bf k},c} a_{{\bf k},c} \rangle$,
and hole 
$f^{h}_{\bf k} \equiv \langle a_{{\bf k},v} a^\dagger_{{\bf k},v}\rangle$
distributions. The vector potential ${\bf A}(z,t) = A(z,t) {\bf e}_E$
determines the electric field
$E(z,t) = - \partial A(z,t) / \partial t$ and its direction
${\bf e}_E$. The corresponding $P$ contains the dipole-matrix element
$d_{\rm{cv}}$ while $J$ depends on the current-matrix element
$j_\lambda(k) = -|e|\hbar {\bf k} \cdot {\bf e}_E /m_\lambda$ 
with the effective mass $m_\lambda$.
If the electric field is in the THz regime, the $A$-dependent part
$J_A \equiv - \frac{1}{\cal S} \sum_{{\bf k},\lambda} e^2 A/m_0 
f^{\lambda}_{\bf k}$ of the intraband current and the off-resonant
interband $P$ contribute only to the refractive index; together, they
provide a Drude-like response if Coulomb interaction is
neglected.\cite{Haug:84,Kira:03}
As a result, {\it $J_{\rm THz} = \sum_{{\bf k},\lambda} j_\lambda({\bf k})  
f^{\lambda}_{\bf k}$ entirely determines the absorption part of the THz response}.

In order to compute
$J_{\rm THz}$ for the fully interacting
system, one has to determine the carrier dynamics 
%-----------------------
%-----------------------
\begin{eqnarray}
   \frac{\partial}{\partial t}
   f^e_{\bf k}
   &=&
   -\frac{2}{\hbar}
   {\rm Im}
   \left[
   P^{\star}_{\bf k}
   \Omega_{\bf k}
   +
     \sum_{{\bf q},{\bf k}'} V_{{\bf k}'+{\bf q}-{\bf k}}
     c^{{\bf q},{\bf k}',{\bf k}}_X
     -
     \sum_{{\bf q},{\bf k}'} V_{{\bf q}}
     c^{{\bf q},{\bf k}',{\bf k}}_{cc}
   \right] + \frac{\partial f^e_{\bf k}}{\partial t} \Biggr|_{\rm ph},
\label{eq:fe_dyn}
\end{eqnarray}
%---------------------
%---------------------
where the last term symbolizes the phonon and photon induced contributions.
A similar equation holds for $f^h_{\bf k}$. Here, 
$\Omega_{\bf k} \equiv d_{\rm{cv}} E(0,t) + 
\sum_{{\bf k}'} V_{{\bf k} - {\bf k}'} P_{{\bf k}'}$
is the renormalized Rabi frequency 
with the Coulomb matrix element $V_{\bf k}$.
The true two-particle correlations are defined by 
$c_{\lambda \lambda'}^{{\bf q},{\bf k}',{\bf k}} \equiv
\Delta 
\langle a^{\dagger}_{{\bf k},\lambda} a^{\dagger}_{{\bf k}',\lambda'}  
a_{{\bf k}'+{\bf q},\lambda}  a_{{\bf k}-{\bf q},\lambda'} \rangle
=
\langle a^{\dagger}_{{\bf k},\lambda} a^{\dagger}_{{\bf k}',\lambda'}  
a_{{\bf k}'+{\bf q},\lambda}  a_{{\bf k}-{\bf q},\lambda'} \rangle
- 
\langle a^{\dagger}_{{\bf k},\lambda} a^{\dagger}_{{\bf k}',\lambda'}  
a_{{\bf k}'+{\bf q},\lambda}  a_{{\bf k}-{\bf q},\lambda'} 
\rangle_{\rm single}$
where the factorized single particle contributions are removed to 
identify excitonic correlation $c_{X} \equiv c_{cv}$ as well as correlated
electrons $c_{cc}$ and holes $c_{vv}$.

In the THz problem, we solve the two-particle quantities
at the same consistent level as the single-particle quantities. 
More specifically, we use the so-called cluster 
expansion to identify all lower level particle clusters in general many-body
expectation values.\cite{Fricke:96,Kira:02a,Hoyer:03} By keeping all terms up 
to two-particle clusters exactly and three-particle clusters at the scattering 
level, the THz dynamics results from
%-----------------------
%-----------------------
\begin{eqnarray}
  i \hbar\frac{\partial}{\partial t} 
  P_{{\bf k}} 
  &=& 
  \tilde{\epsilon}_{{\bf k}} 
  P_{{\bf k}} 
+ j({\bf k}) A(0,t) P_{\bf k} 
- \left[1-f^e_{{\bf k}}-f^h_{{\bf k}}\right] \Omega_{{\bf k}} 
+\frac{\partial P_{\bf k}}{\partial t} \Biggr|_{\rm ph},
\nonumber\\
&+&
  \sum_{{\bf k}',{\bf l},\lambda} V_{{\bf l}}
\left[
    \Delta \langle 
           a^{\dagger}_{{\bf k},v} a_{{\bf k}',\lambda}^{\dagger}
           a_{{\bf k}'+{\bf l},\lambda}
           a_{{\bf k}-{\bf l},c}
           \rangle
-
    \Delta \langle
           a_{{\bf k},c}^{\dagger} a_{{\bf k}',\lambda}^{\dagger}
           a_{{\bf k}'+{\bf l},\lambda}
           a_{{\bf k}-{\bf l},v}
           \rangle^{\star}
\right],
\label{eq:SBE-Pol}
\\
%%%%%%%%%%%%%%%%%%%%%%%%%%%%%%%%%%%%%%%%%%%%%
%%%%%%%%%%%%%%%%%%%%%%%%%%%%%%%%%%%%%%%%%%%%%
  i \hbar\frac{\partial}{\partial t}
  c^{{\bf q},{\bf k}',{\bf k}}_{\rm X}
&=& \Delta E^{{\bf q},{\bf k}',{\bf k}} c^{{\bf q},{\bf k}',{\bf k}}_{\rm X}
+ S^{{\bf q},{\bf k}',{\bf k}}
+j({\bf k}'+{\bf q}-{\bf k}) A(0,t) 
   c_{\rm X}^{{\bf q},{\bf k}',{\bf k}} 
%\nonumber\\
%  &+&
  +
  \left(
     1-f^e_{{\bf k}}-f^h_{{\bf k}-{\bf q}}
  \right)
  \sum_{\bf l}
	V_{{\bf l}-{\bf k}}
   c^{{\bf q},{\bf k}',{\bf l}}_{\rm X}
\nonumber\\
&-&
  \left(
     1-f^e_{{\bf k}'+{\bf q}}-f^h_{{\bf k}'}
  \right) 
  \sum_{\bf l}
      V_{{\bf l}-{\bf k}'}
      c^{{\bf q},{\bf l},{\bf k}}_{\rm X}
%\nonumber\\
%&+&
+
D^{{\bf q},{\bf k}',{\bf k}}_{\rm rest}
+\gamma^{{\bf q},{\bf k}',{\bf k}}_{{\bf l},{\bf m},{\bf n}}
 c_{\rm X}^{{\bf l},{\bf m},{\bf n}}
\label{eq:EXPcvcv},
\end{eqnarray}
%-----------------------
%-----------------------
with the phonon and photon induced correlations 
$\frac{\partial P_{\bf k}}{\partial t}|_{\rm ph}$.
The carrier-carrier correlations
$c_{cc}$ and $c_{vv}$ obey a similar equation as $c_{X}$,
however, they do not include the direct THz source proportional to
$j({\bf k})\,A(0,t) \equiv [j_e({\bf k}) + j_h({\bf k})]A(0,t)$.
The different contributions to the $c_X$ 
dynamics contain the renormalized kinetic energy
$\Delta E^{{\bf q},{\bf k}',{\bf k}}$ and the scattering source $S$
which has the typical Coulomb scattering form 
$ V \left[
f_{1} f_{2} \left(1-f_{3}\right) \left(1-f_{4}\right)
- \left(1-f_{1}\right) \left(1-f_{2}\right) f_{3} f_{4}\right]$.
The explicit summation over $V$ and $c_X$
includes the attractive interaction between correlated electron-hole
pairs allowing for the possibility of bound excitons. 
The remaining two-particle Coulomb and photon contributions
are denoted as $D_{\rm rest}$ and the term proportional to $\gamma$ 
symbolizes the three-particle Coulomb and phonon terms; 
all these terms solved in our numerical 
approach. Equations (\ref{eq:fe_dyn})--({\ref{eq:EXPcvcv}) together with 
Maxwell's wave equation present a general description of optical interband as
well as intraband THz excitations. When only optical interband fields are
applied, the $j A$-terms can be omitted
and Eqs.~(\ref{eq:fe_dyn})--(\ref{eq:SBE-Pol}) 
reduce to the semiconductor Bloch equations while
Eq.~(\ref{eq:EXPcvcv}) describes the formation dynamics of incoherent 
excitons.\cite{Haug:94,Kira:01,Hoyer:03}

In the absence of THz fields, all carrier quantities have even symmetry 
with respect to a sign change of 
${\bf k}$, e.g.~$f^{\lambda}_{-{\bf k}} = f^{\lambda}_{{\bf k}}$
while $j_\lambda({\bf k})$ is odd such that
$J_{\rm THz} = \sum_{{\bf k},\lambda} 
j_\lambda({\bf k}) f^{\lambda}_{\bf k} = 0$.
Finite THz response follows when
we subdivide all terms into odd and even quantities via
$f^{\lambda}_{{\bf k},O(E)} \equiv 
[f^{\lambda}_{{\bf k}}
\;{\rm {}^{\;-}_{(+)}}\;
f^{\lambda}_{-{\bf k}}]/2$,
$P_{{\bf k},O(E)} \equiv [P_{{\bf k}}  
\;{\rm {}^{\;-}_{(+)}}\; P_{-{\bf k}}]/2$,
and
$c^{{\bf q},{\bf k}',{\bf k}}_{\lambda \lambda',O(E)} \equiv 
[c^{{\bf q},{\bf k}',{\bf k}}_{\lambda \lambda'}  
 \;{\rm {}^{\;-}_{(+)}}\;
c^{-{\bf q},-{\bf k}',-{\bf k}}_{\lambda \lambda'}]/2$. Inserting
this general decomposition into 
Eqs.~(\ref{eq:SBE-Pol}) and (\ref{eq:EXPcvcv}), 
we find that, e.g., 
an odd $P_{O}$ is driven by $j\,A\,P_{E}$  
and an odd  $c_{X,O}$ follows from a $j\,A\,c_{X,E}$ where $j$
provides the needed symmetry breaking. Once $P_{O}$ 
or $c_{X,O}$ are generated, also densities with odd symmetry
are created via Eq.~(\ref{eq:fe_dyn}) and
$J_{\rm THz} = \sum_{{\bf k},\lambda} j_\lambda({\bf k})  
f^{\lambda}_{{\bf k},O}$ produces a nonvanishing THz response.

In this Letter, we concentrate on the time-resolved 
linear THz response of a system that has been excited by an optical 
pulse resonant with band-to-band transitions. The THz field is then applied 
at different time delays with respect to the optical excitation. 
The corresponding THz dynamics follows from the odd
parts of Eqs.~(\ref{eq:fe_dyn})--({\ref{eq:EXPcvcv}).
Besides  $J_{\rm THz}$, also $c_{cc,O}$ and $c_{vv,O}$ are indirectly 
generated, which leads to the decay of $J_{\rm THz}$.
When $J_{\rm THz}$ is computed from the dynamics of 
$f^{\lambda}_{{\bf k},O}$, we notice that the coherent $P_{{\bf k},O}$ 
and the incoherent $c_{X,O}$ occur as additive sources in 
Eq.~(\ref{eq:fe_dyn}). The THz response from a coherent polarization 
has already been studied within the framework of
the semiconductor Bloch equations.\cite{Meier:94,Hughes:00}
However, when the interband coherences vanish, the nature of the
$J_{\rm THz}$ changes since 
{\it in the incoherent regime the linear THz response results 
exclusively from correlated electron-hole pairs}.

In the computations, the single- and two-particle dynamics of
Eqs.~(\ref{eq:fe_dyn})--({\ref{eq:EXPcvcv}) contain the bare 
Coulomb-matrix element while screening originates 
microscopically from the coupling to the higher order clusters.
The three-particle scattering terms $\gamma$ are approximated
to have a statically screened Coulomb interaction formally resulting from
coupling to four-particle clusters.\cite{Kira:02a,Jahnke:97}
In the following, we compute the THz response of a QW
and of an array of quantum wires. Because of the demanding numerics, 
the fully dynamic solution of Eqs.~(\ref{eq:fe_dyn})--(\ref{eq:EXPcvcv})
is evaluated for the quantum wire system;
the QW system is solved with for a postulated configuration of excitonic
correlations. In both cases, we assume GaAs type parameters; the 3D-Bohr
radius is $a_0=12.4$\,nm and the wire radius is chosen such that the energy
separation between the two lowest exciton states is 5~meV. The corresponding
8~nm wide QW has similar excitonic states. 
At low temperatures and relatively small interband
pump excess energies studied here,
it is sufficient to include only acoustic phonons\cite{Thraen:00}.
The specific forms of the Coulomb and phonon-matrix elements 
and scattering are discussed in Ref.~\cite{Kira:02a,Hoyer:03}.

We first apply our
THz theory to monitor the build up of correlations after a nonresonant
optical pulse excitation where a rapid screening of the bare Coulomb interaction 
can be expected.\cite{Banyai:98,Kwong:00} 
Typically, the screened interaction 
$V^{S}_{\bf q} = V_{\bf q}/\epsilon_{\rm L}(\omega,{\bf q})$ 
is defined by the longitudinal dielectric function
$\epsilon_{\rm L}(\omega,{\bf q})$ with frequency and 
momentum dependence. In the mean field approximation,
$\epsilon_{\rm L}(\omega,{\bf q}) = 1 - \omega^2_{\rm PL}({\bf q})/\omega^2$
such that
$1/\epsilon_{\rm L}$ has a resonance at the plasmon frequency 
$\omega_{\rm PL}({\bf q})$.
%Recent experiments\cite{Huber:01} suggest that
%the THz response can detect this resonance.
The corresponding transversal
dielectric function follows from 
$\epsilon_{\rm T}(\omega) = 1 + \chi_{\rm T}(\omega)$ with 
$\chi_{\rm T}(\omega) = 
i J_{\rm THz}(\omega)/[\omega E_{\rm THz}(\omega)]$.

\begin{figure}[h]
\resizebox{0.3\textwidth}{!}{%
\includegraphics{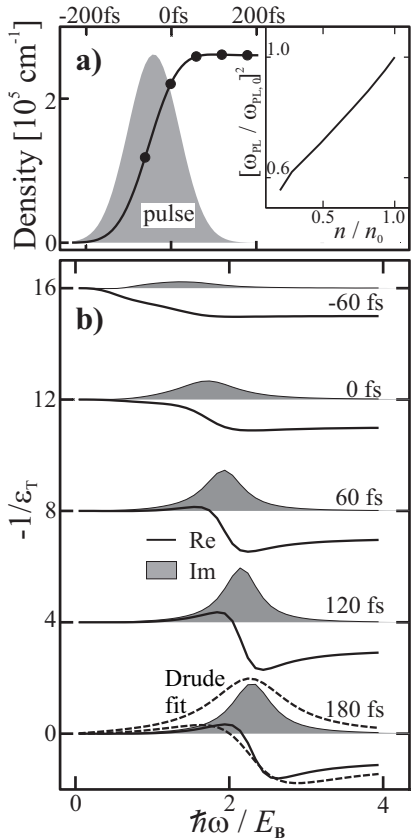}}
\caption{
(a) Time evolution of carrier density (solid line) and 
exciting pulse $|E|^2$ (shaded area). 
(b) Real (solid line) and imaginary part (shaded area) of
$-1/\epsilon_T$ for different THz probe delays
(solid circles in frame above) as function of THz energy ($E_B = 4.2$\,meV). 
Inset shows density dependence of $-1/\epsilon_T$ resonance; 
$n_0$ and $\omega_{\rm PL,0}$ correspond to frame (b).
}
\label{fig1}
\end{figure}
Figure \ref{fig1} shows the computed time-evolution of $1/\epsilon_{\rm T}$
for a quantum-wire system that has been excited into the carrier continuum
21\,meV above the excitonic resonance with an ultrafast pulse 
excitation. According to Fig.~\ref{fig1}a, a quasi stationary carrier 
density is built up very quickly and we choose $t = 0$ to correspond
to almost fully developed density.
Following
Ref.~\cite{Huber:01}, we plot $1/\epsilon_{\rm T}(\omega)$ in Fig.~\ref{fig1}b
for the different delay times indicated by the circles in Fig.~\ref{fig1}a.
For comparison we plot as dashed line a simple Drude
response fit, $\chi_{\rm fit} (\omega) = 
- \omega^{2}_{\rm PL}/[\omega (\omega + i \delta)]$.
Our calculations show that the
more or less structureless early time response develops into
a clear $1/\epsilon_{\rm T}(\omega)$ resonance 
after about 60~fs.
This resonance narrows and shifts toward its final position 
within 180~fs. These dynamical features are very similar to the
experimental observations\cite{Huber:01} and the 180~fs
response follows from the dynamic build up of Coulomb 
correlations. The inset to Fig.~\ref{fig1}a shows that the computed
resonance frequency is proportional to the square root of the density, 
confirming the basic feature of a plasmon resonance.\cite{Haug:94}

At first sight, the observation of a finite plasmon resonance 
in a QW or quantum-wire system is surprising since a postulated
relation $\epsilon_{\rm T}(\omega) = \epsilon_{\rm L}(\omega,{\bf 0}) = 
1 - \omega^2_{\rm PL}({\bf 0})/\omega^2$\cite{photon_mom}
predicts $\epsilon_{\rm T}(\omega) = 1$ because 
$\omega_{\rm PL}({\bf 0})$ vanishes in one- and 
two-dimensions.\cite{Haug:94}
To obtain some insights into the origin of our 
numerically computed response,
we study $\chi_{\rm T}(\omega)$ when
the Coulomb and phonon sums in Eq.~(\ref{eq:EXPcvcv}) are omitted.
In this case, $c_{\rm X,O}$ can 
be solved analytically and the result, inserted into 
Eq.~(\ref{eq:fe_dyn}), shows that 
a part of $S^{{\bf q},{\bf k}',{\bf k}}$ provides
Lindhard-like screening contributions to $J_{\rm{THz}}$. 
We then obtain
$\chi_{\rm T}(\omega) = - \omega^{-2} \sum_{\bf q} 
\omega_{\rm PL}^2({\bf q}) S_{\bf q} = 
-\omega_{\rm PL}^2({\bf q}_{\rm eff})/\omega^{2}$
where the Coulomb scattering kernel $S_{\bf q}$
leads to nonvanishing effective ${\bf q}_{\rm eff}$
and $\omega_{\rm PL}({\bf q}_{\rm eff})$. Hence, we identify 
the Coulomb scattering as relevant source providing 
a build up of the finite plasmon response.

\begin{figure}[h]
\resizebox{0.3\textwidth}{!}{%
\includegraphics{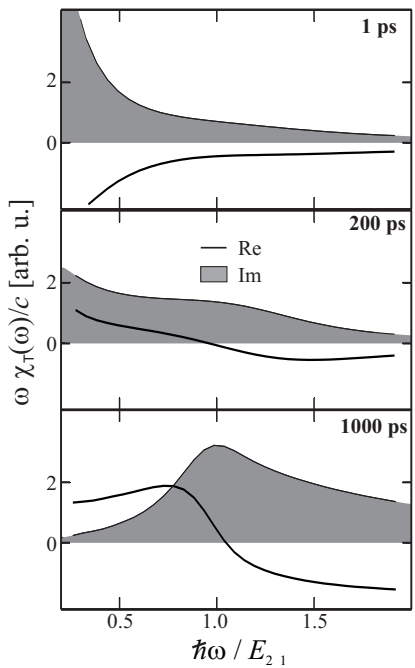}}
\caption{
Terahertz absorption (shaded area) and refractive index change
(solid line)  for different THz probe delays after nonresonant 
excitation. Here, $E_{21} = 5$\,meV and final density is 
$6 \times 10^4\;{\mathrm{cm}}^{-1}$.
}
\label{fig2}
\end{figure}
In comparison to $1/\epsilon_{\rm T}$, plain $\epsilon_{\rm T}$
may display resonances related to exciton populations.\cite{Kira:01}
Following recent experiments \cite{Kaindl:03}, we compute
THz absorption spectrum
$\alpha(\omega) = \omega {\rm Im}\left[\chi_{\rm T}(\omega)\right]/c$.
As in Fig.~\ref{fig1}, the system is 
nonresonantly excited but we assume a lower excitation intensity and
focus on the long time limit to obtain conditions favorable to
excitons. Figure \ref{fig2} shows that the 
computed $\alpha(\omega)$ is very broad and shows 
no resonances at 1~ps after the excitation.\cite{plasmon}
Even after 200~ps, the THz 
response has changed only moderately due to the slow phonon scattering 
from electron-hole plasma to excitons. However, roughly 1~ns 
after the excitation, $\alpha(\omega)$ shows a pronounced 
resonance at 5~meV corresponding exactly to the 
energy difference between the two lowest exciton states. At the same time,
$\omega {\rm Re}\left[\chi_{\rm T}(\omega)\right]/c$
becomes positive for low frequencies in contrast to the 1~ps-response. 
The asymmetric shape of $\alpha(\omega)$ results from transitions 
between the lowest and all other exciton states. These results are in
good agreement with recent experiments.\cite{Kaindl:03}

\begin{figure}[h]
\resizebox{0.3\textwidth}{!}{%
\includegraphics{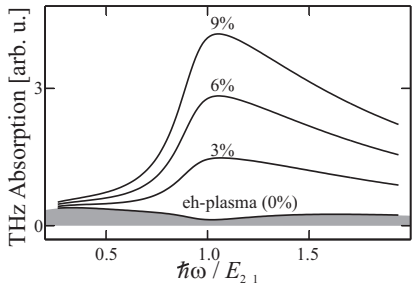}}
\caption{
Terahertz absorption of QW with different 
postulated exciton fractions.
Carrier system at 40~K with  density 
$4 \times 10^9\;{\mathrm{cm}}^{-2}$, and 
$E_{21} = 7$\,meV.
}
\label{fig3}
\end{figure}
In order to evaluate the THz response of a two-dimensional QW system, we
make use of the results published previously.\cite{Hoyer:03}
There, a careful analysis of the exciton formation after a band-to-band excitation
revealed that the many-body state $c_{\rm X,E}$ corresponding to the 
late times of Fig.~\ref{fig2} can be accurately described as a 
superposition of a correlated plasma and a
time dependent 1s-exciton population. In Fig.~\ref{fig3} we display the
results for such a postulated superposition using
different 1s-exciton concentrations. 
The comparison with Fig.~\ref{fig2} shows that the THz response 
is qualitatively independent of the dimensions.
Furthermore, we observe that the correlated plasma absorption is 
very much suppressed compared to the result expected from a 
simplified Drude analysis. Consequently, the transitions from
the 1s to the 2p exciton dominate the absorption
even for small exciton concentrations.

In summary, a microscopic analysis of the THz response leads to 
a symmetry breaking where a transversal THz field generates 
odd quantities with respect to the carrier momentum.
In the incoherent regime, the THz response follows entirely from the 
correlated electron-hole pairs and can sensitively detect the
presence of exciton populations. The computations also demonstrate that
the exciton resonance emerges slowly after an off-resonant excitation
while the plasmon resonance is built up on an ultrafast time scale 
even in low-dimensional systems.

\begin{acknowledgments}
We thank D.S.Chemla for stimulating discussions about THz experiments.
This work was supported by the Deutsche Forschungsgemeinschaft through
the Quantum Optics in Semiconductors Research Group,
by the Humboldt Foundation and the Max-Planck Society through the
Max-Planck Research prize, and by the Optodynamics Center of the
Philipps-Universit{\"a}t Marburg.
\end{acknowledgments}


\begin{thebibliography}{99}
%
\bibitem{Groeneveld:94}
R.M.~Groeneveld and D.~Grischkowsky,
J.~Opt.~Soc.~Am.~B {\bf 11}, 2502 (1994).
%
\bibitem{Cerne:96}
J. Cerne {\it et al.},
Phys.~Rev.~Lett.~{\bf 77}, 1131 (1996).
%
\bibitem{Beard:00}
M.C.~Beard, G.M.~Turner, and C.A.~Schmuttenmaer,
Phys.~Rev.~B {\bf 62}, 15764 (2000)

\bibitem{Huber:01}
R.~Huber {\it et al.},
Nature~{\textbf{414}}, 286--289 (2001).
%
\bibitem{Kaindl:03}
R.A.~Kaindl, M.A.~Carnahan, D.~H{\"a}gele, R.~L{\"o}venich, and D.S.~Chemla,
accepted for publication in Nature (2003).
%
\bibitem{Haug:94}
H.~Haug and S.W.~Koch,
{\it Quantum Theory of the Optical and Electronic Properties of Semiconductors}
(World Scientific Publ., Singapore, 1994).
%
\bibitem{Cohen:89}
C.~Cohen-Tannoudji, J.~Dupont-Roc, and G.~Grynberg,
{\it Photons \& Atoms}, (Wiley, New York, 1989).
%
\bibitem{Haug:84}
H.~Haug and S.~Schmitt-Rink,
Prog.~Quantum Electron.~{\bf 9}, 3 (1984).
%
\bibitem{Kira:03}
M.~Kira, W.~Hoyer, and S.W.~Koch,
Phys.~Stat.~Sol.~(b), accepted.
%
\bibitem{Fricke:96}
J.~Fricke, Ann.~Phys.~{\bf 252}, 479 (1996).
%
\bibitem{Kira:02a}
M.~Kira, W.~Hoyer, and S.W.~Koch,
J.~Nonlin.~Opt.~B~{\bf 29}, 481 (2002).
%
\bibitem{Hoyer:03}
W. Hoyer, M. Kira, and S. W. Koch,
Phys.~Rev~B~{\bf 15}, 155113 (2003).
%
\bibitem{Kira:01}
M.~Kira {\it et al.},
Phys.~Rev.~Lett.~\textbf{87}, 176401 (2001).
%
\bibitem{Meier:94}
T.~Meier {\it et al.},
Phys.~Rev.~Lett. \textbf{73}, 902 (1994).
%
\bibitem{Hughes:00}
S.~Hughes and D.~Citrin,
J.~Opt.~Soc.~Am.~B \textbf{17}, 128 (2000).
%
\bibitem{Jahnke:97}
F.~Jahnke, M.~Kira, and S.W.~Koch
Z.~Physik B \textbf{104}, 559 (1997).
%
\bibitem{Thraen:00}
A.~Thr{\"a}nhardt {\it et al.},
Phys.~Rev.~B \textbf{62}, 2706 (2000).
%
\bibitem{Banyai:98}
L.~B\'{a}nyai {\it et al.},
Phys.~Rev.~Lett.~{\bf 81}, 882 (1998).
%
\bibitem{Kwong:00}
N.-H.~Kwong and M.~Bonitz,
Phys.~Rev.~Lett.~{\bf 84}, 1768 (2000).
%
\bibitem{plasmon}
The corresponding
$1/\epsilon_{\rm T}$ does show a plasmon resonance in agreement
with Fig.~\ref{fig1}. 
%
\bibitem{photon_mom}
The momentum of THz photons is practically vanishing at the scale 
typical for carriers.
%
\end{thebibliography}
\end{document}